\documentclass[twocolumn,showpacs,preprintnumbers,amsmath,amssymb,nofootinbib,aps,prc,10pt]{revtex4-1}
\usepackage{graphicx}
\usepackage{dcolumn}
\usepackage{bm}

\usepackage{pstricks}
\usepackage{psboxit}
\usepackage{color}
\usepackage{hyperref}

\makeatletter
\def\HyPsd@CatcodeWarning#1{}
\makeatother

\begin{document}
\newcommand{\D}{{\rm d}}

\title{Single-particle dissipation in TDHF studied from a phase-space perspective}
\author{N. Loebl$^1$, A. S. Umar$^2$, J.A. Maruhn$^1$, P.-G. Reinhard$^3$, P.D. Stevenson$^4$, and V. E., Oberacker$^2$}
 
\affiliation{
$^1$Institut fuer Theoretische Physik, 
Universitaet Frankfurt, D-60438 Frankfurt, Germany 
} 
\affiliation{
$^2$Department of Physics and Astronomy, Vanderbilt University,
Nashville, Tennessee 37235, USA
}
\affiliation{
$^3$Institut fuer Theoretische Physik II,
Universitaet Erlangen-Nuernberg, D-91058 Erlangen, Germany
}
\affiliation{
$^4$Department of Physics, University of Surrey, Guildford, GU2 7XH, UK
}

\date{\today}

\begin{abstract}
We study dissipation and relaxation processes within the 
time-dependent Hartree-Fock approach using the Wigner distribution function.
On the technical side we present a geometrically unrestricted
framework which allows us to calculate the full six-dimensional Wigner
distribution function. With the removal of geometrical constraints, we
are now able to extend our previous phase-space analysis of heavy-ion
collisions in the reaction plane to unrestricted mean-field simulations of nuclear matter on a
three-dimensional Cartesian lattice. From the physical point of view we provide a quantitative analysis
on the stopping power in TDHF. This is linked to the effect of transparency. For the medium-heavy
$^{40}$Ca+$^{40}$Ca system we examine the impact of different
parametrizations of the Skyrme force, energy-dependence, and the significance of extra time-odd terms
in the Skyrme functional.
\end{abstract}


\maketitle

\section{Introduction}

Time-dependent Hartree-Fock (TDHF) theory provides a fully self-consistent mean-field approach to nuclear dynamics.
First employed in the late 1970's~\cite{Bonche,Svenne,Negele,Davies} the
applicability of TDHF was constrained by the limited
computational power. Therefore, early applications 
treated the problem in only one spatial dimension, utilizing a very
simplified parametrization of the nuclear interaction.
Due to the increase in computational power, state-of-the-art
TDHF calculations are now feasible in three-dimensional
coordinate space, without any symmetry restrictions and using the full Skyrme 
interaction~\cite{Kim,Simenel,Nakatsukasa,Umar05a,Maruhn1,Guo08a}.

In this work the Wigner distribution function~\cite{Wigner} is
calculated as an analysis tool to probe the phase space behavior in
TDHF evolution of nuclear dynamics. 
In comparison to previous work~\cite{Loebl}, where the Wigner 
analysis was performed in one and two dimensions,
we are now able to carry out both the TDHF simulation
and the phase-space analysis in three dimensions. 
Transformation from 
coordinate-space representation to phase-space representation, i.e. calculating the Wigner
distribution from the density matrix, still remains a computationally challenging problem.
Here, we
present a fully three-dimensional analysis which allows the study of
relaxation processes simultaneously in all directions in $k$-space.
An early one-dimensional study of the Wigner function for
TDHF can be found in~\cite{Maruhn3}.  

The paper is outlined as follows: In Sec.~\ref{sec:wigner} we
introduce the Wigner distribution function and discuss the
numerical framework used in this work. We then
introduce the principal observables summarizing the local or global momentum-space properties
of the Wigner function. First the quadrupole operator in momentum space 
which gives rise to the usual deformation parameters $\beta$ and $\gamma$ to
probe relaxation processes in dynamical calculations.
In addition, we define an estimate for the occupied phase-space volume to
obtain a relation between the fragment separation in momentum and
coordinate space. 
 
Sec.~\ref{sec:results} provides a detailed discussion of the
central $^{40}$Ca+$^{40}$Ca collision, paying
particular attention to the effect of transparency. We discuss
the impact of different Skyrme parametrizations on the 
relaxation behavior, as well as the dependence on the center-of-mass energy
for a fixed Skyrme interaction. We also examine the influence of extra
time-odd terms in the Skyrme functional. 

\section{Outline of formalism}
\label{sec:wigner}

\subsection{Solution of the TDHF equations}

The TDHF equations are solved on a three-dimensional Cartesian lattice
with a typical mesh spacing of $1$\:fm. The initial
setup of a dynamic calculation needs a static Hartree-Fock run, whereby the
stationary ground states of the two fragments are computed with the
damped-gradient iteration algorithm~\cite{Blum,Reinhard}. 
The TDHF runs are initialized with energies above the Coulomb barrier at some
large but finite separation. The two ions are boosted with velocities obtained by assuming
that the two nuclei arrive at this initial separation on a Coulomb trajectory.
The time propagation is managed by
utilizing a Taylor-series expansion of the time-evolution operator~\cite{Flocard}
up to sixth order with a time step of $t=0.2$\:fm/c.
The spatial derivatives are calculated using the fast Fourier transforms (FFT).

\subsection{Computing the Wigner function}

The Wigner distribution function is obtained by a partial Fourier
transform of the density matrix
$\rho(\mathbf{r}\!-\!\frac{\mathbf{s}}{2},    
\mathbf{r}\!+\!\frac{\mathbf{s}}{2},t)$, with respect to the 
relative coordinate $\mathbf{s}=\mathbf{r}-\mathbf{r}'$

\begin{eqnarray}
 f^{(3)}_\mathrm{W}(\mathbf{r},\mathbf{k},t)
 &=&
  \int\frac{{\rm d}^3 s}{(2\pi)^3}\:
  e^{-i\mathbf{k}\cdot\mathbf{s}} 
  \rho(\mathbf{r}\!-\!\frac{\mathbf{s}}{2},
       \mathbf{r}\!+\!\frac{\mathbf{s}}{2},t)
  \;,
\\
  \rho(\mathbf{r},\mathbf{r}',t)
  &=&
  \sum_{l}\Psi^\dagger_{l}(\mathbf{r},t)\Psi_{l}(\mathbf{r}',t)
\:.
\end{eqnarray}
Because $f_\mathrm{W}$ is not
positive definite, it is misleading to consider the Wigner function as a phase-space probability
distribution. We will refer to the appearance of negative
values for $f_\mathrm{W}$ in Sec.~\ref{sec:results}.

Evaluating the Wigner function in six-dimensional phase space is still a computational  challenge and only possible 
employing full Open MP parallelization and extensive use of FFT's. The determing factor is the grid size, which results in
\begin{equation}
N_{x}^{2}\log{(N_x)}\star N_{y}^{2}\log{(N_y)}\star
N_{z}^{2}\log{(N_z) }
\end{equation}
steps to provide the Wigner transform in full
space, where $N_x,N_y,N_z$ are the grid points in each direction. 
Storing the Wigner function reduced to the reaction
plane, i.e. $f^{(3)}_\mathrm{W}(x,y=0,z,\mathbf{k})$ at one time step will consume
$\sim 140$\;Mb of disk space for the applications presented in Sec.~\ref{sec:results}. Going to larger grid sizes, 
needed for heavier systems, and/or storing the full three-dimensional Wigner function will clearly result in entering
the Gb regime. 

\subsection{Observables}
\label{sec:observ}

In this section we discuss some of the observables used in our analysis.
In order to avoid any misunderstandings we will label all observables evaluated in momentum space with
a subscript $k$, and all observables in coordinate space with a
subscript $r$.

\subsubsection{Quadrupole in momentum space}

As an observable to probe relaxation in phase-space quantitatively, we
evaluate the quadrupole operator in momentum space. The local
deviation of the momentum distribution from a spherical shape is a
direct measure for equilibration. The local quadrupole tensor in
$k$-space is given by
\begin{equation}
Q^{ij}_k(\mathbf{r},t)=\int \D^3 k
\left[ 3\langle k_i(\mathbf{r},t) \rangle \langle k_j(\mathbf{r},t)
\rangle - \langle \mathbf{k}^2(\mathbf{r},t) \rangle
\delta_{ij}\right]\:,
\end{equation}
using the $m$-th moment from the local momentum
distribution 
\begin{equation}
  \langle\mathbf{k}^{(m)}(\mathbf{r},t)\rangle
  = 
  \frac{\int {\rm d}^3 k \:(\mathbf{k}-\langle\mathbf{k}(\mathbf{r},
          t)\rangle)^{m}f^{(3)}_\mathrm{W}(\mathbf{r},\mathbf{k},t)}
       {\int {\rm d}^3
k\:f^{(3)}_\mathrm{W}(\mathbf{r},\mathbf{k},t)}
  \:,
\end{equation}
with  $\langle\mathbf{k}(\mathbf{r}, t)\rangle$ denoting the average
local flow
\begin{equation}
  \langle\mathbf{k}(\mathbf{r},t)\rangle
  = 
  \frac{\int {\rm d}^3
k\:\mathbf{k}\:f^{(3)}_\mathrm{W}(\mathbf{r},\mathbf{k},t)}
       {\int {\rm d}^3
k\:f^{(3)}_\mathrm{W}(\mathbf{r},\mathbf{k},t)}\:.
\end{equation}
The spherical quadrupole moments $Q_k^{20}(\mathbf{r},t)$ and
$Q_k^{22}(\mathbf{r},t)$ are computed by diagonalization of
$Q_k^{ij}(\mathbf{r},t)$
\begin{eqnarray}
\label{eq:diag}
Q_k^{20}(\mathbf{r},t)&=&\sqrt{\frac{5}{16\pi}}\lambda_3  \\
Q_k^{22}(\mathbf{r},t)&=&\sqrt{\frac{5}{96\pi}}(\lambda_2 - \lambda_1)
\end{eqnarray}
with $\lambda_3 > \lambda_2 > \lambda_1$ labeling the eigenvalues of
$Q_k^{ij}(\mathbf{r},t)$. Switching to polar notation the
observables
\begin{eqnarray}
\beta_k(\mathbf{r},t)&=&
\sqrt{\beta_{20}^2(\mathbf{r},t)+2\beta_{22}^2(\mathbf{r},t)} \\
\gamma_k(\mathbf{r},t)&=&
|\arctan{\frac{\sqrt{2}\beta_{22}(\mathbf{r},t)}{\beta_{20}(\mathbf{r}
,t)}}\frac{ 180^{\circ}
}{\pi}|\: ,
\end{eqnarray}
are obtained via the dimensionless quantities
\begin{eqnarray}
\beta_k^{20}(\mathbf{r},t)&=&\frac{4\pi Q_{20}}{5 r_k^2
\rho(\mathbf{r},t)} \\
\beta_k^{22}(\mathbf{r},t)&=&\frac{4\pi Q_{22}}{5 r_k^2
\rho(\mathbf{r},t)}\:,
\end{eqnarray}
where
\begin{equation}
r_{k}(\mathbf{r},t)=\sqrt{\langle\mathbf{k}(\mathbf{r},
t)\rangle^2/\rho(\mathbf{r}, t) }\: ,
\end{equation}
accounts for the local rms-radius in $k$-space. The norm is defined such that
\begin{equation}
\rho(\mathbf{r},t)=\int \D^3
k\:f^{(3)}_\mathrm{W}(\mathbf{r},\mathbf{k},t)\:.
\end{equation}
In the presented formalism it is straightforward to define global
observables. The global quadrupole tensor is calculated by
spatial integration
\begin{equation}
Q_k^{ij}(t)=\int \D^3 r\: 
\rho(\mathbf{r},t)Q^{ij}_k(\mathbf{r},t)\:.
\end{equation}
Applying the same diagonalization as in the local case (\ref{eq:diag})
we end up with a global definition for $\beta_k^{20}(t)$ and
$\beta_k^{22}(t)$. For the following results we exclusively use the
global definition since it is more compact and allows the
simultaneous visualization of multiple time-dependent observables. 

\subsubsection{Quadrupole in coordinate space}

To illustrate the global development of a reaction, we will also use
the expectation value $Q^{20}_r\equiv\langle \hat{Q}^{20}_r\rangle$ of
the quadrupole operator in coordinate space.

\subsubsection{Occupied phase space volume}

To give a rough measure for the phase-space volume occupied by
the fragments during a heavy-ion collision
we assume a spherical shape of the local momentum distribution. Adding
up the $k$-spheres 
\begin{equation}
V_k(\mathbf{r,t})=\frac{4\pi}{3}\langle\mathbf{k}^2(\mathbf{r},
t)\rangle^{3/2}\: ,
\end{equation}
leads to the total occupied phase-space volume
\begin{equation}
V_k(t)=\int {\rm d}^3 r \: V_k(\mathbf{r,t})\:.
\end{equation}

\section{Results and discussion}
\label{sec:results}

It is the aim of this work to provide a quantitative analysis of the 
magnitude of relaxation processes occurring in TDHF. Therefore we will
vary a single reaction parameter, while all the other parameters are
fixed. The $^{40}$Ca+$^{40}$Ca-system provides a suitable test
case. This particular choice is motivated by Ref.~\cite{Sim11} where the
applicability of TDHF was demonstrated up to very high energies. All calculations
in this section were done for central collisions (impact parameter $b=0$).
The numerical grid was set up with $36\times24^2$ grid points.

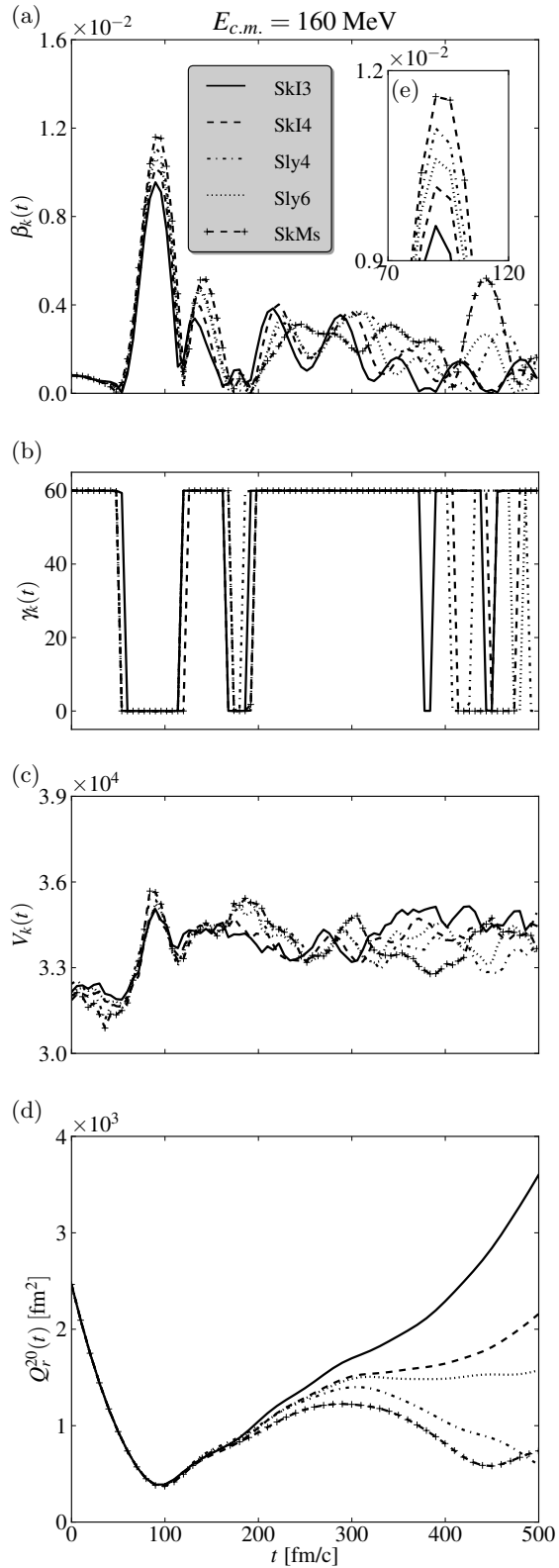
\begin{figure}[hbtp]
\centering
  \newsavebox\IBox
  \savebox\IBox{\includegraphics[width=8.1cm]{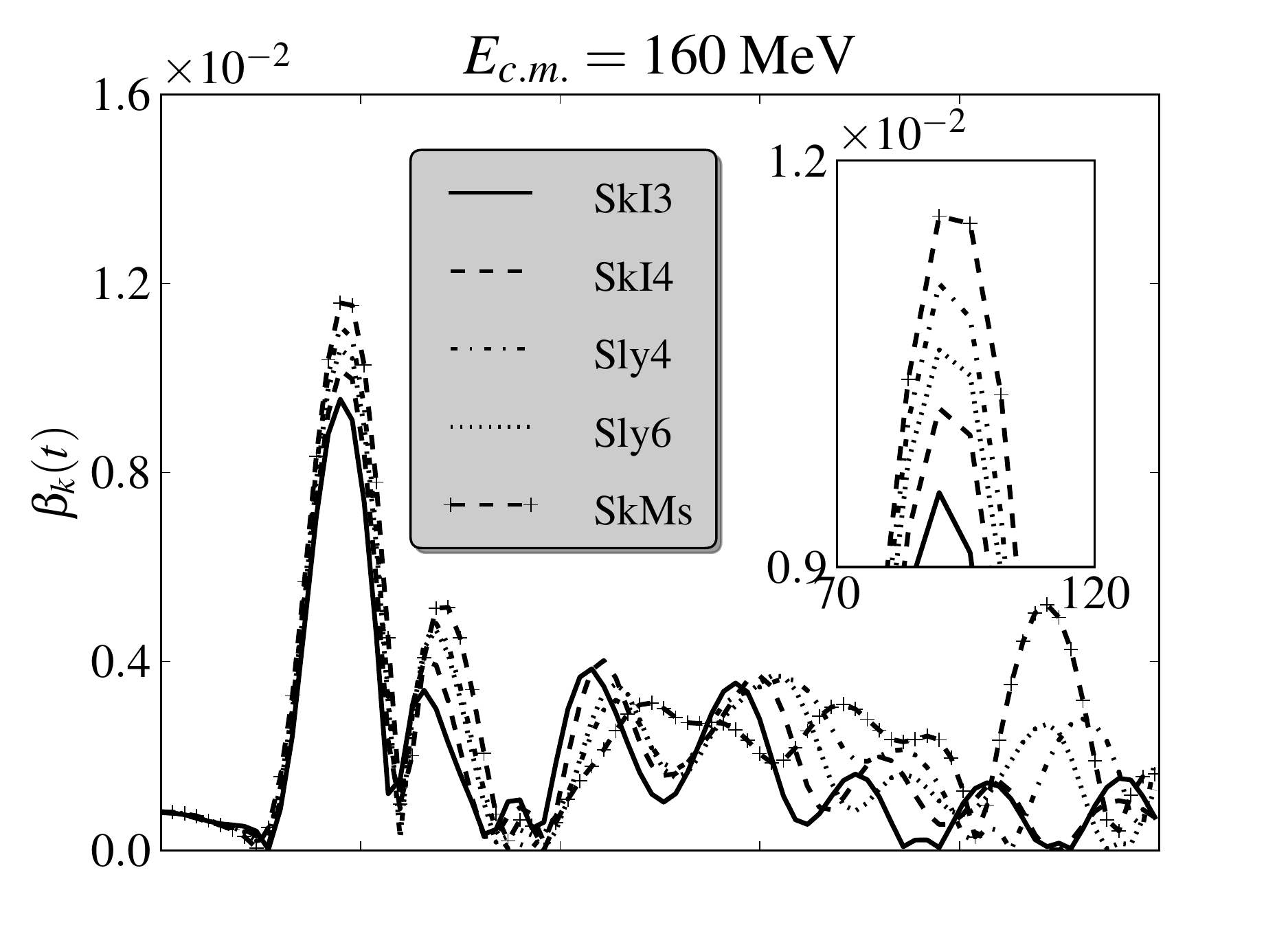}}
 \begin{pspicture}(\wd\IBox,\ht\IBox)
    \rput[lb](-0.5,0.5){\usebox\IBox}
    \rput[lb](-0.3,6.){(a)}  
    \rput[lb](4.8,5.){(e)}  
  \end{pspicture}
  \savebox\IBox{\includegraphics[width=8.1cm]{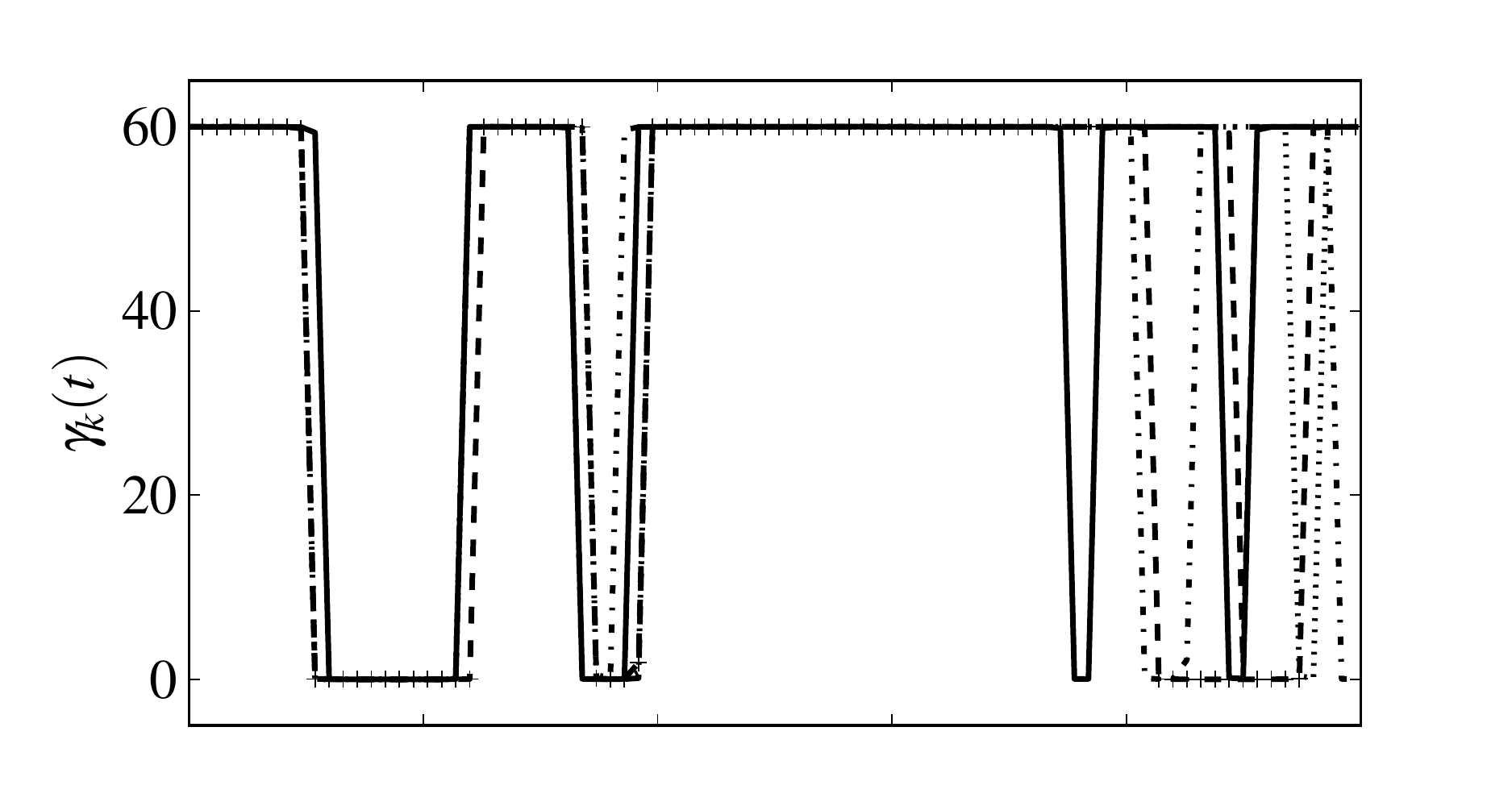}}
 \begin{pspicture}(\wd\IBox,\ht\IBox)
    \rput[lb](-0.5,0.5){\usebox\IBox}
    \rput[lb](-0.3,4.5){(b)}  
  \end{pspicture}
  \savebox\IBox{\includegraphics[width=8.1cm]{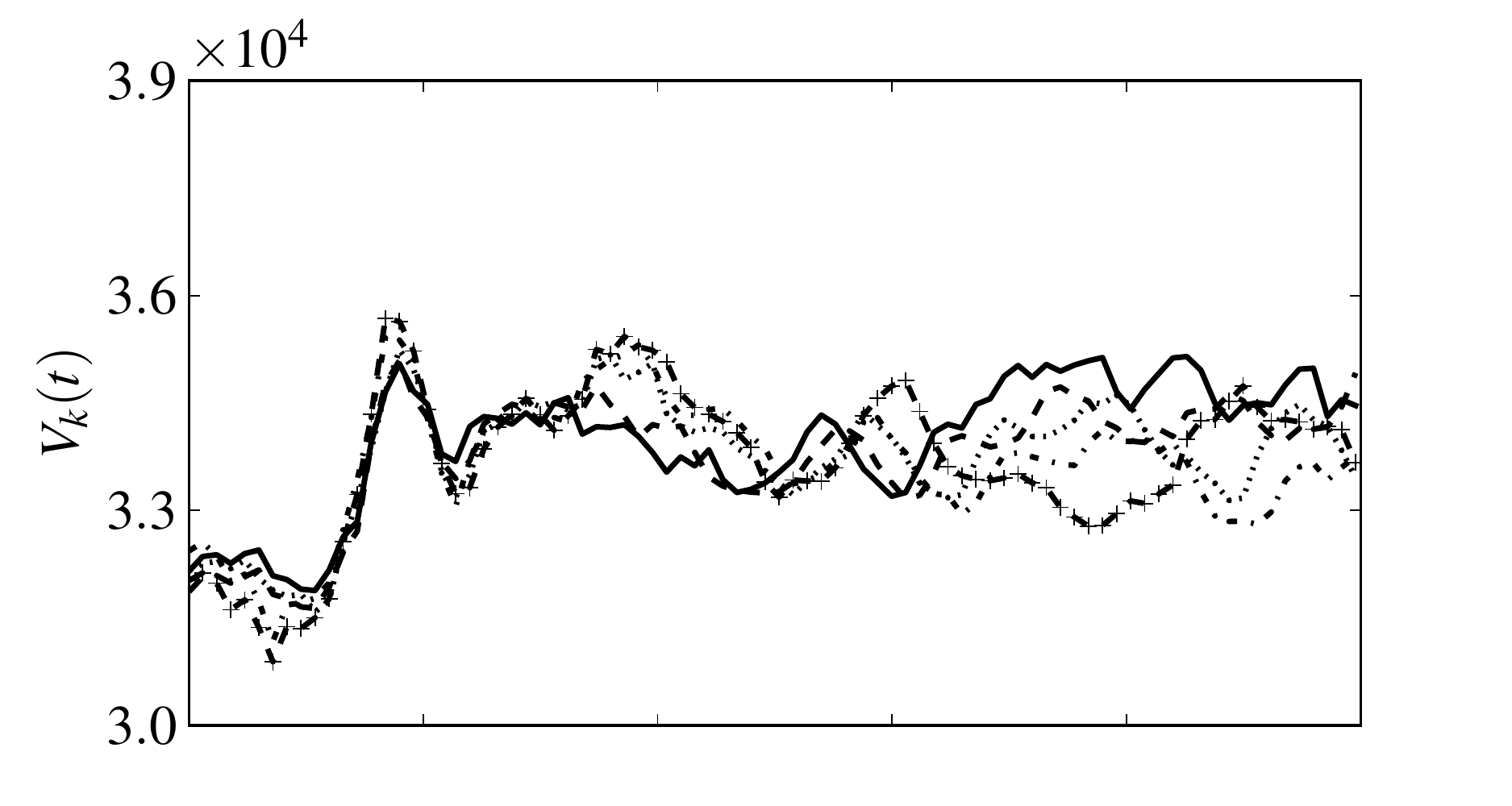}}
 \begin{pspicture}(\wd\IBox,\ht\IBox)
    \rput[lb](-0.5,0.5){\usebox\IBox}
    \rput[lb](-0.3,4.5){(c)}  
  \end{pspicture}
  \savebox\IBox{\includegraphics[width=8.1cm]{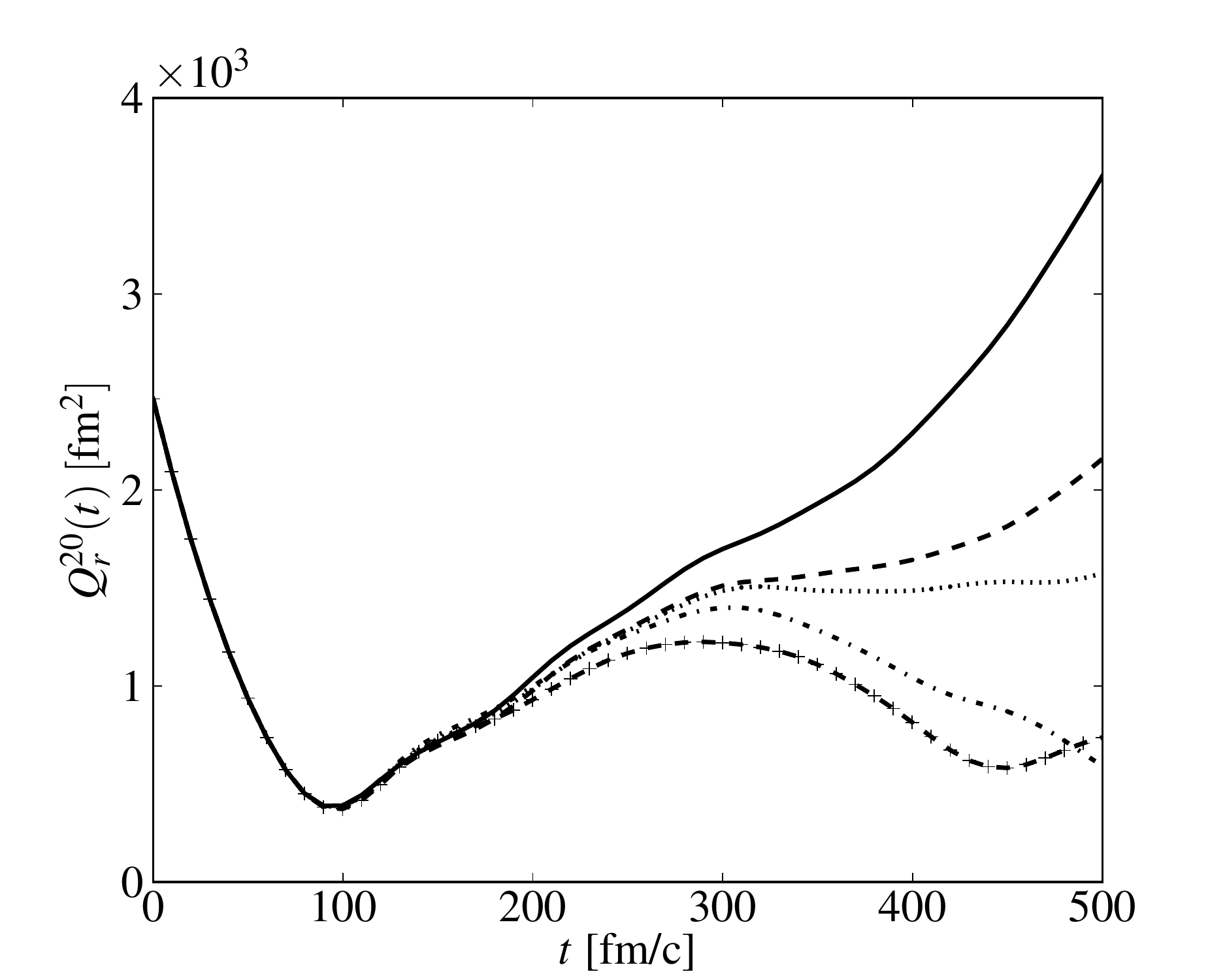}}
 \begin{pspicture}(\wd\IBox,\ht\IBox)
    \rput[lb](-0.5,0.5){\usebox\IBox}
    \rput[lb](-0.3,6.5){(d)}  
  \end{pspicture}
\caption{\label{fig:Skyrme}
Global observables $\beta_k(t)$ (a), $\gamma_k(t)$ (b), $V_k(t)$ (c),
and $Q^{20}_r(t)$ (d) are shown for a central $^{40}$Ca+$^{40}$Ca
collision with a center-of-mass energy $E_\mathrm{c.m.}=160\;$MeV. Each curve
corresponds to a different Skyrme force as indicated in the legend.}
\end{figure}

\subsection{Variation of the Skyrme force}

In the first set of calculations we vary the Skyrme parametrization.
Figure~\ref{fig:Skyrme} shows the results of a central
$^{40}$Ca+$^{40}$Ca collision with the Skyrme parametrizations SLy4,
SLy6 \cite{Chabanat}, SkMs \cite{Bartel}, SkI3, and SkI4 \cite{Reinhard2}.
While SkMs was chosen as an example for an outdated
interaction, the SLy(X) set of forces was originally
developed to study isotopic trends in neutron rich nuclei and neutron
matter with applications in astrophysics. The SkI(X) forces take the
freedom of an isovector spin-orbit force into account. This results in
an improved description of isotopic shifts of r.m.s. radii in
neutron-rich Pb isotopes.

The global development of the reaction is visualized in subplot (d).
The time-dependent expectation value $Q^{20}_r(t)$ shows the five
trajectories initially in good agreement but finally fanning out. A
similar splitting behavior 
depending on the employed Skyrme parametrization was already found in
\cite{Maruhn85}. While the two Sk(X)-forces
show a full separation of the two fragments, there is a slight
remaining contact between the fragments for the case of SLy6, which
will result in complete
separation in a longer calculation. However the trajectories for SLy4
and SkMs show a merged system in the final state, which was found to
persist in long-time simulations.

We now consider the relationship between the observed characteristics in coordinate space
with the dynamics in phase space. Subplot (a) shows the
$\beta_k$-value, measuring the global deviation of the momentum
distribution from a sphere. The initial
$\beta_k$-peak is strongly damped for all five Skyrme-forces. While
the time development for all parametrizations
remain in phase up to the second peak, later it starts to vary and 
continue with damped oscillations.
For a better visualization the first peak is magnified in subplot (e). The taller
the $\beta_k$-peak the longer the fragments will stick together in coordinate space. The 
effect appears to depend on the effective mass $m^*/m$.  Smaller effective masses give rise to a smaller $\beta_k$-deformation. Table \ref{tab:effectm}
summarizes the $m^*/m$-values associated with the
maximal deformations $\beta_k^{max}$ for all the
Skyrme forces used in this work. It seems logical that
the $m^*/m$-dependence is visible in the phase-space analysis since it is directly linked to
the nucleons' kinetic motion. It is harder to randomize the directed motion of a nucleon with
a higher effective mass than it is for a nucleon with a smaller $m^*/m$-dependence.

\begin{table}
\caption{\label{tab:effectm}
Effective masses of Skyrme parametrizations used in this work are
listed in connection with the maximal $\beta_k$-values from the plots
(a, e) in Fig.~\ref{fig:Skyrme}.}
\begin{ruledtabular}
\begin{tabular}{lll}
Skyrme force &$m^*/m$& $\beta_k^{max}$\\ 
\hline
SkM∗ &0.79&0.0116 \\
SLy4 &0.70&0.0111 \\
SLy6 &0.69&0.0106 \\
SkI4 &0.65&0.0102 \\
SkI3 &0.57&0.0095 \\
\end{tabular}
\end{ruledtabular}
\end{table}
Subplot (b) shows the $\gamma_k$-value which indicates, whether a
deformation is prolate, oblate, or triaxial \cite{Greiner}.
For the present scenario of a central collision the $\gamma_k$-value
jumps between prolate and oblate configurations indicating that the
momentum distribution oscillates between being aligned primarily in
the beam direction or transverse to it. For the sake of completeness
we additionally present the occupied
phase-space volume (c) which will prove more useful for the next
reaction parameter to be discussed: the center-of-mass energy.

\subsection{Variation with the center-of-mass energy}

\begin{figure}[hbtp]
 \centering
  \savebox\IBox{\includegraphics[width=8.1cm]{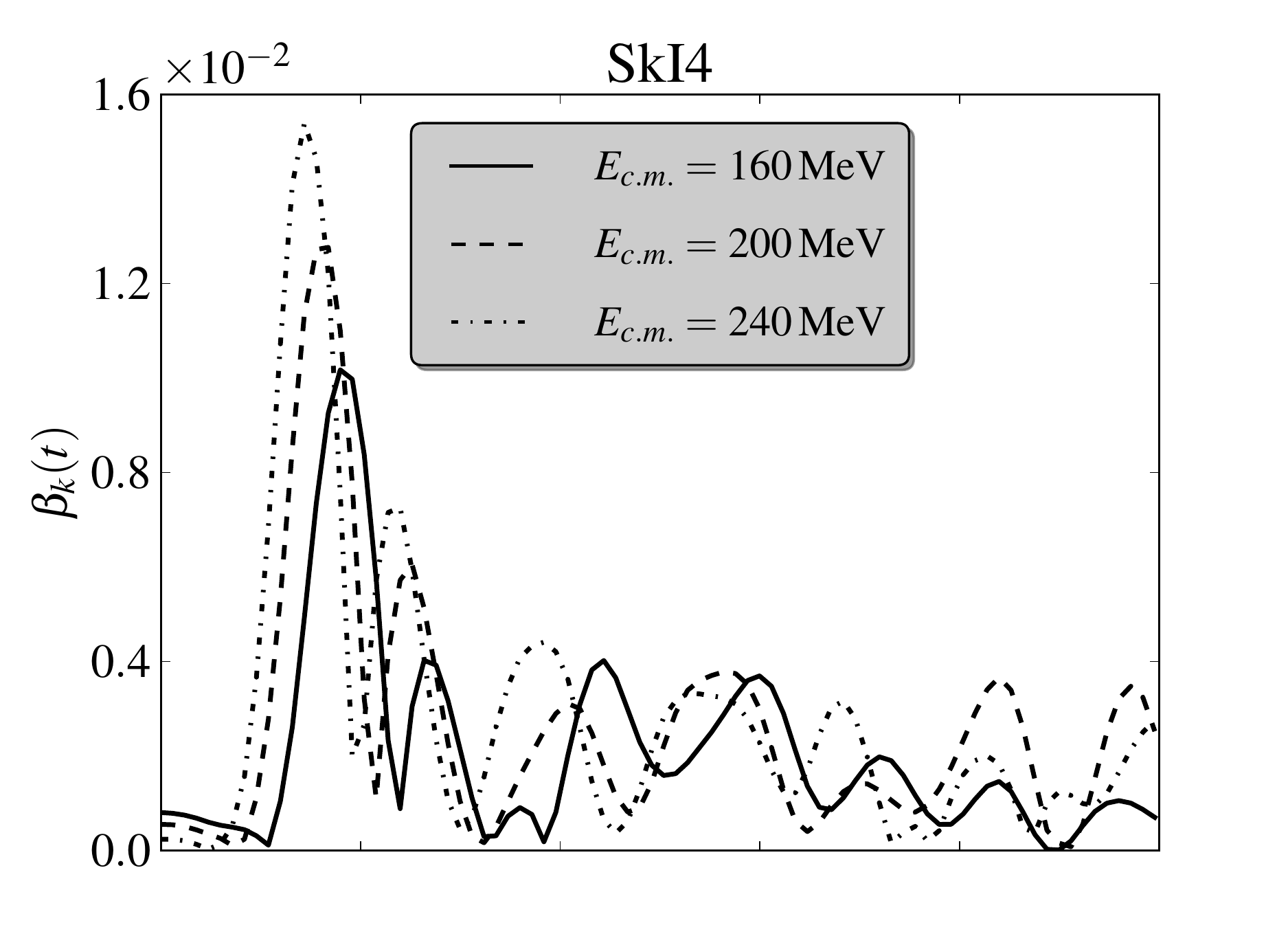}}
 \begin{pspicture}(\wd\IBox,\ht\IBox)
    \rput[lb](-0.5,0.5){\usebox\IBox}
    \rput[lb](-0.3,6.){(a)}  
  \end{pspicture}
  \savebox\IBox{\includegraphics[width=8.1cm]{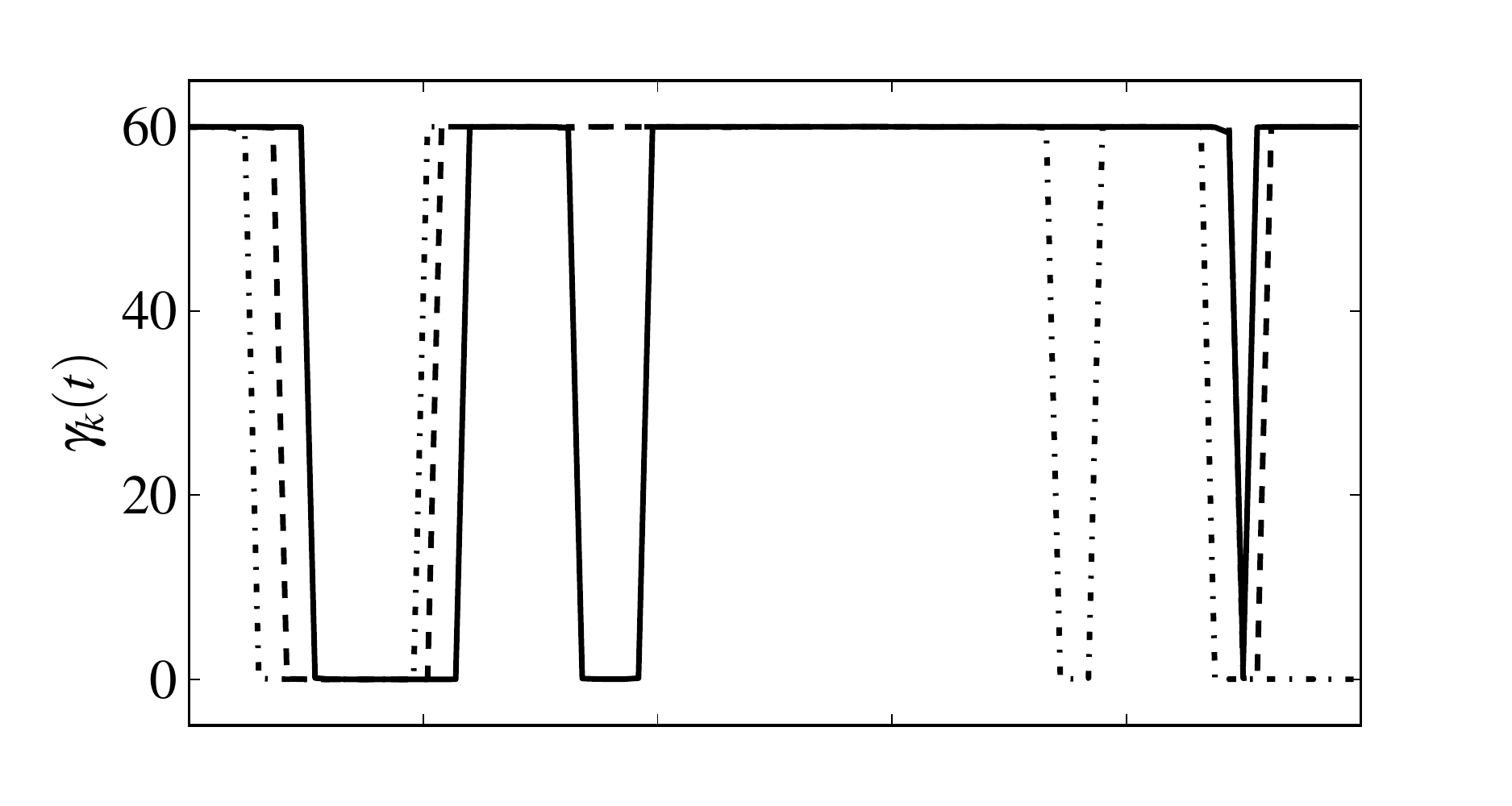}}
 \begin{pspicture}(\wd\IBox,\ht\IBox)
    \rput[lb](-0.5,0.5){\usebox\IBox}
    \rput[lb](-0.3,4.5){(b)}    
  \end{pspicture}
  \savebox\IBox{\includegraphics[width=8.1cm]{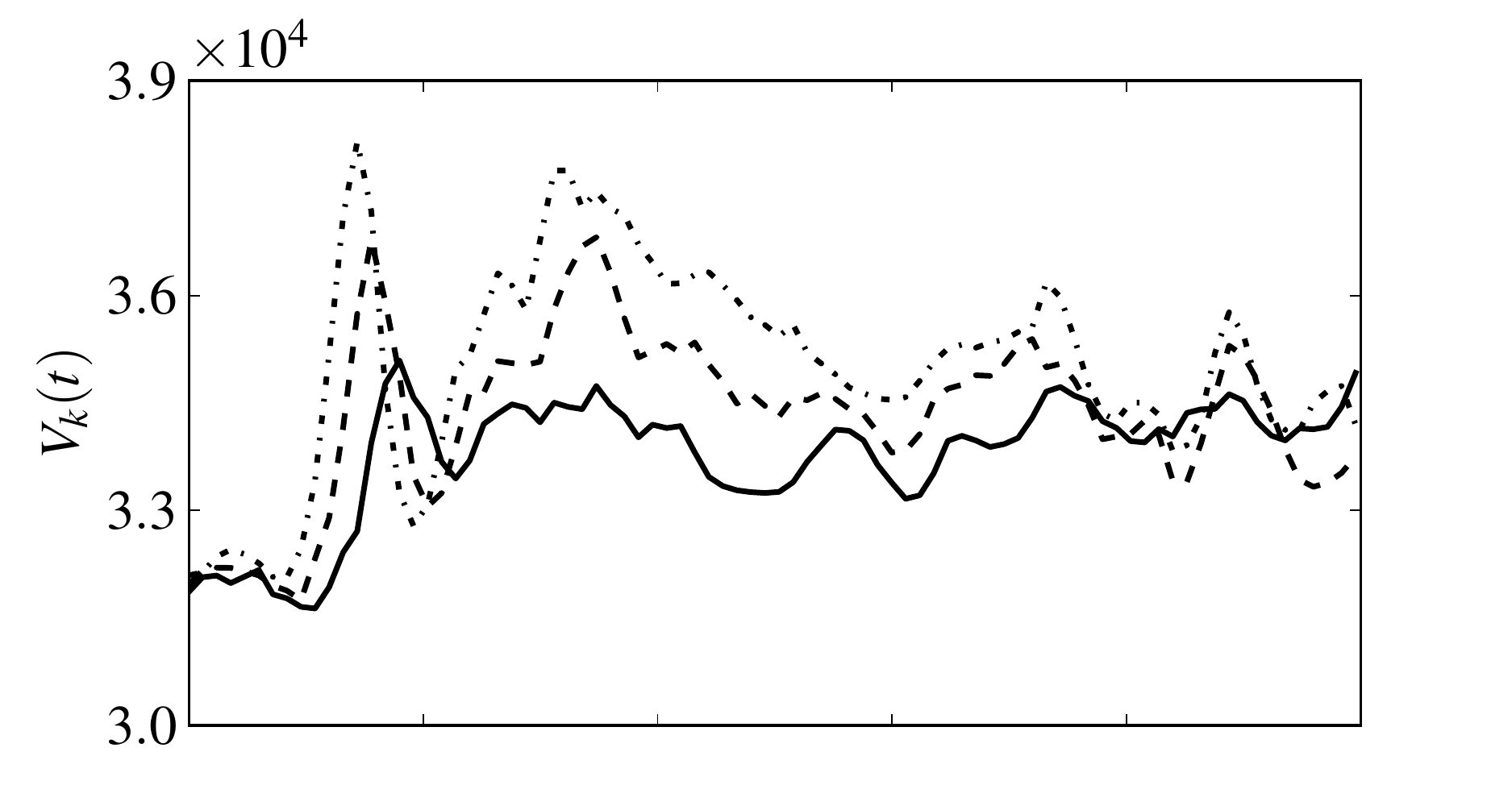}}
 \begin{pspicture}(\wd\IBox,\ht\IBox)
    \rput[lb](-0.5,0.5){\usebox\IBox}
     \rput[lb](-0.3,4.5){(c)}  
  \end{pspicture}
  \savebox\IBox{\includegraphics[width=8.1cm]{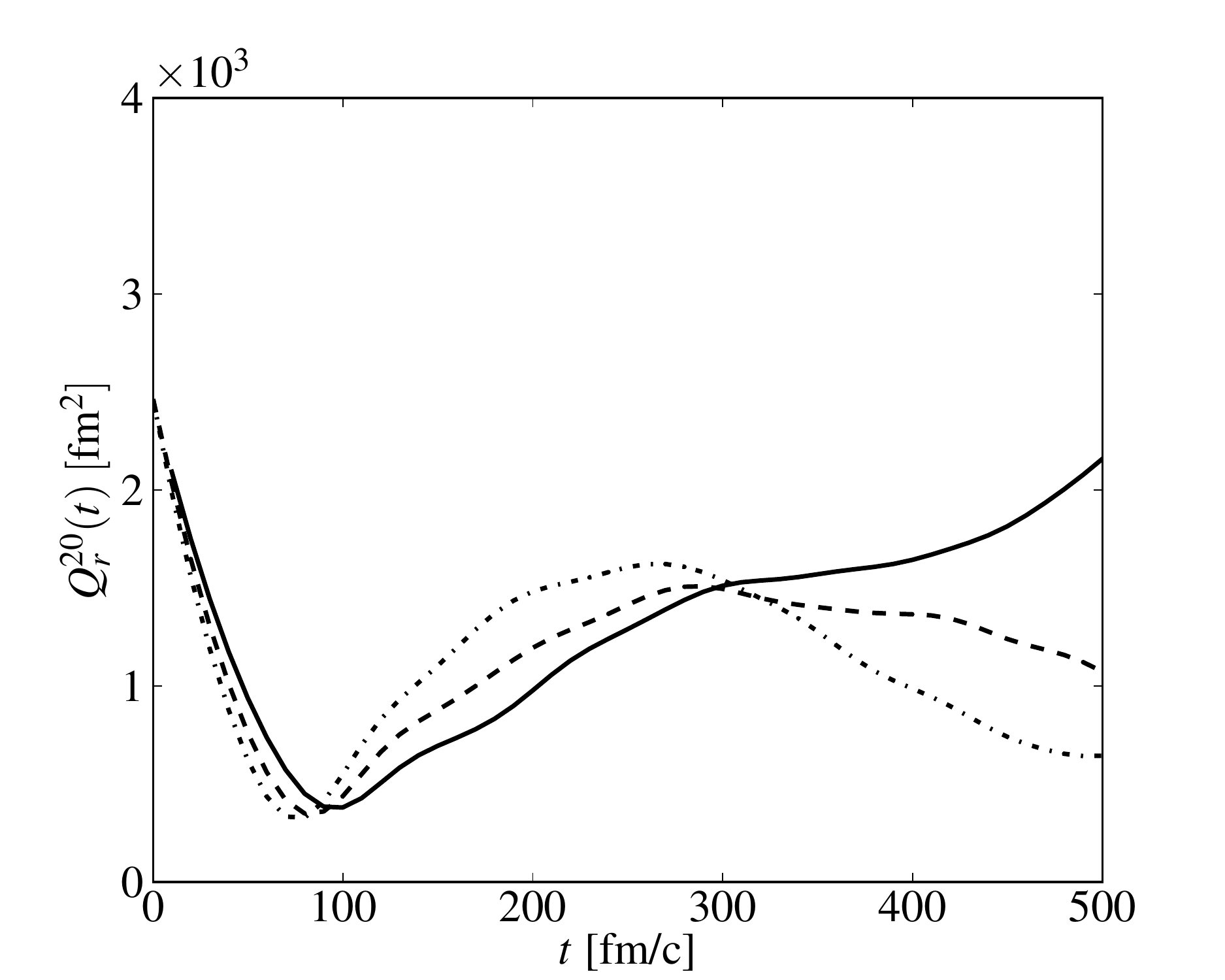}}
 \begin{pspicture}(\wd\IBox,\ht\IBox)
    \rput[lb](-0.5,0.5){\usebox\IBox}
    \rput[lb](-0.3,6.5){(d)} 
  \end{pspicture}
\caption{\label{fig:ecm} 
Global observables $\beta_k(t)$ (a), $\gamma_k(t)$ (b), $V_k(t)$ (c),
and $Q^{20}_r(t)$ (d) are shown for a central $^{40}$Ca+$^{40}$Ca
collision with fixed Skyrme interaction SkI4. Each curve corresponds
to a different center-of-mass energy.}
\end{figure}
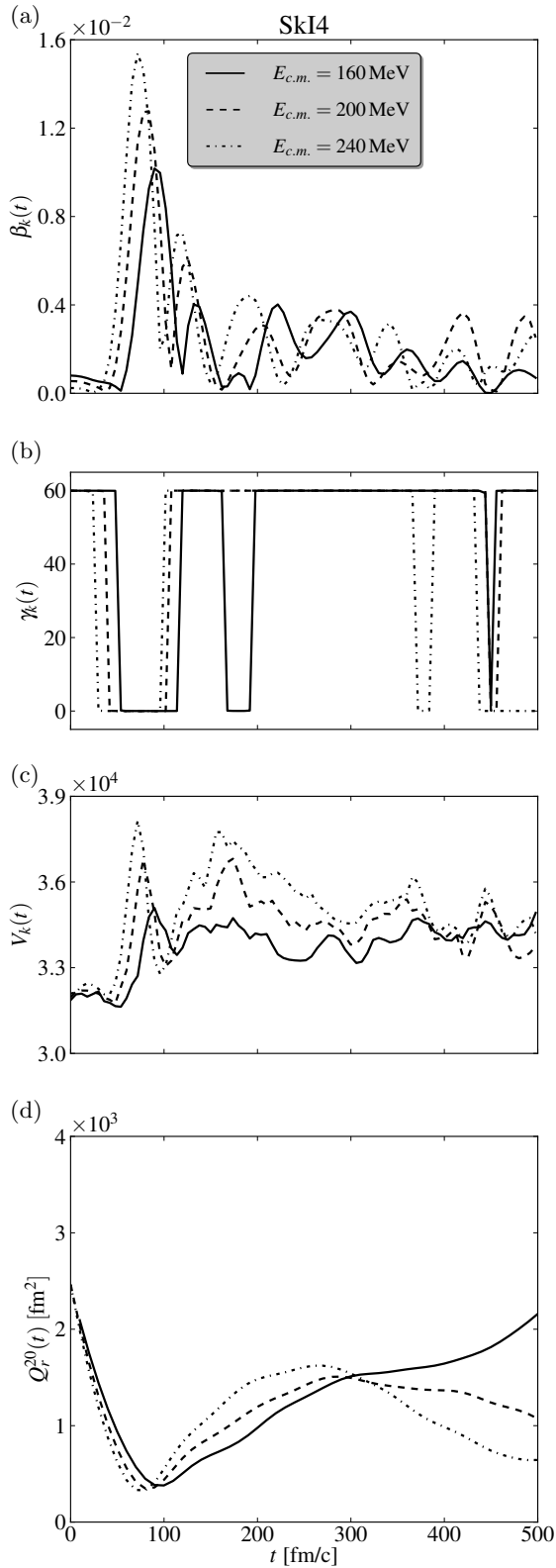

As a second reaction parameter the center-of-mass energy, $E_\mathrm{c.m.}$, is
varied. Results are presented for energies ranging from
$E_\mathrm{c.m.}=2$\:MeV/nucleon up to $E_\mathrm{c.m.}=3$\:MeV/nucleon. The Skyrme
interaction now is fixed to be SkI4. For the case of the lowest
(highest) energy Video~\ref{vid:ca160} (Video~\ref{vid:ca240})
provides a video visualizing the reaction in phase space.  The
calculation done with the lowest energy $E_\mathrm{c.m.}=160$\:MeV
shows two fully separated fragments in the exit channel. In contrast,
the case with the highest energy (as well as the one at an
intermediate energy) results in a merged system. The global
observables are plotted in Fig.~\ref{fig:ecm}.
It may not be obvious at first why the fragments should split for
lower energies and merge for higher ones. But the estimate for the
occupied phase-space volume $V_k(t)$ presented in subplot (c)
indicates that $V$ increases with energy. Therefore the fragments'
average distance in phase space is larger, while in compensation they can come
closer to each other in coordinate space. However, this behavior is also
dependent on the particular Skyrme force used and the presence of time-odd
terms, which is discussed in the next subsection.

\begin{video}
\includegraphics[width=8.1cm]{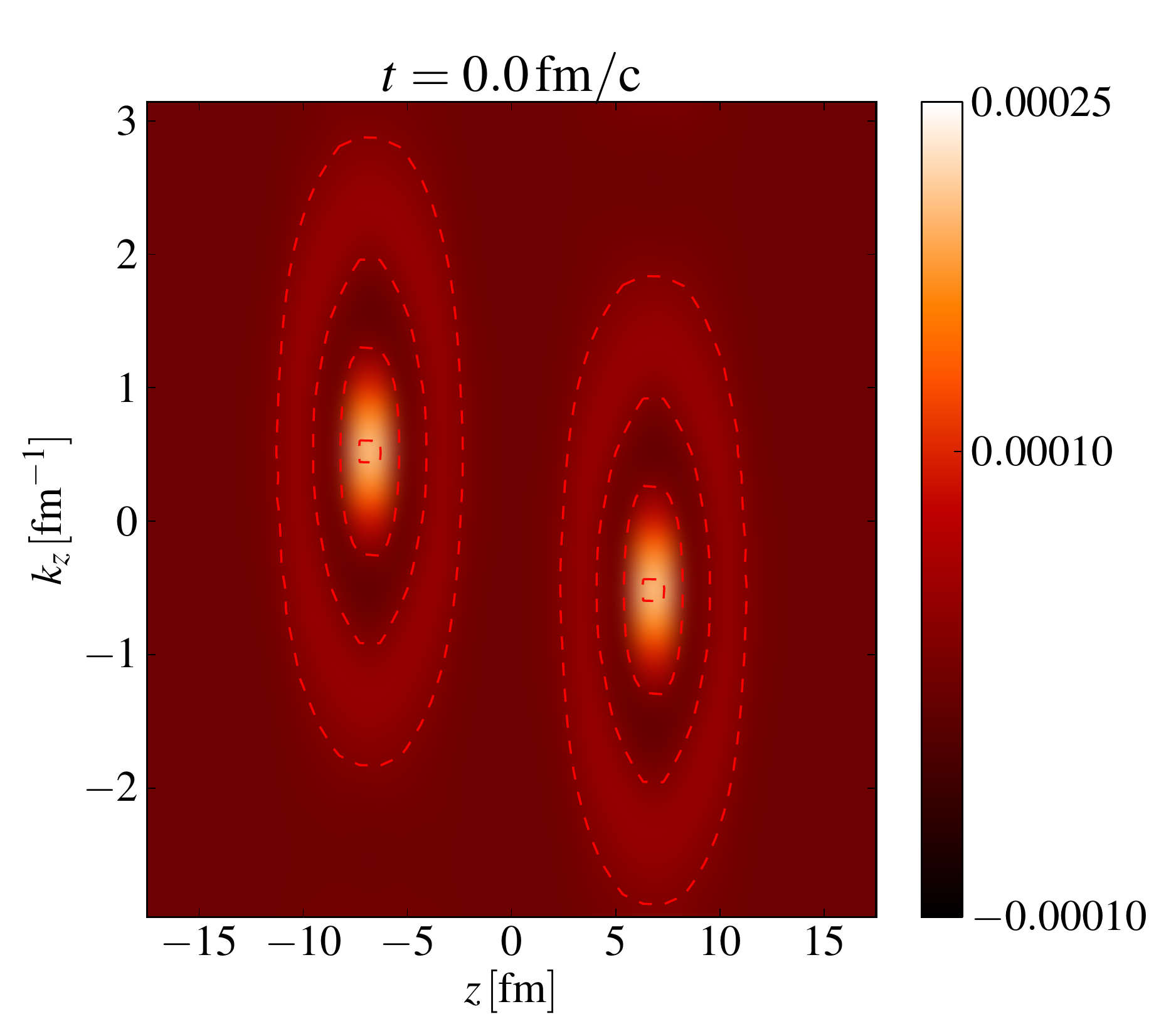}
\setfloatlink{http://th.physik.uni-frankfurt.de/~loebl/vid1.mpeg}
\caption{\label{vid:ca160} (color online)
Two-dimensional $z$-$k_z$-slice from the full six-dimensional Wigner distribution
$f^{(3)}_\mathrm{W}(\mathbf{r},\mathbf{k})$ for a central $^{40}$Ca+$^{40}$Ca collision 
with a center-of-mass energy of $E_\mathrm{c.m.}=160$\:MeV.}
\end{video}
\begin{video}
\includegraphics[width=8.1cm]{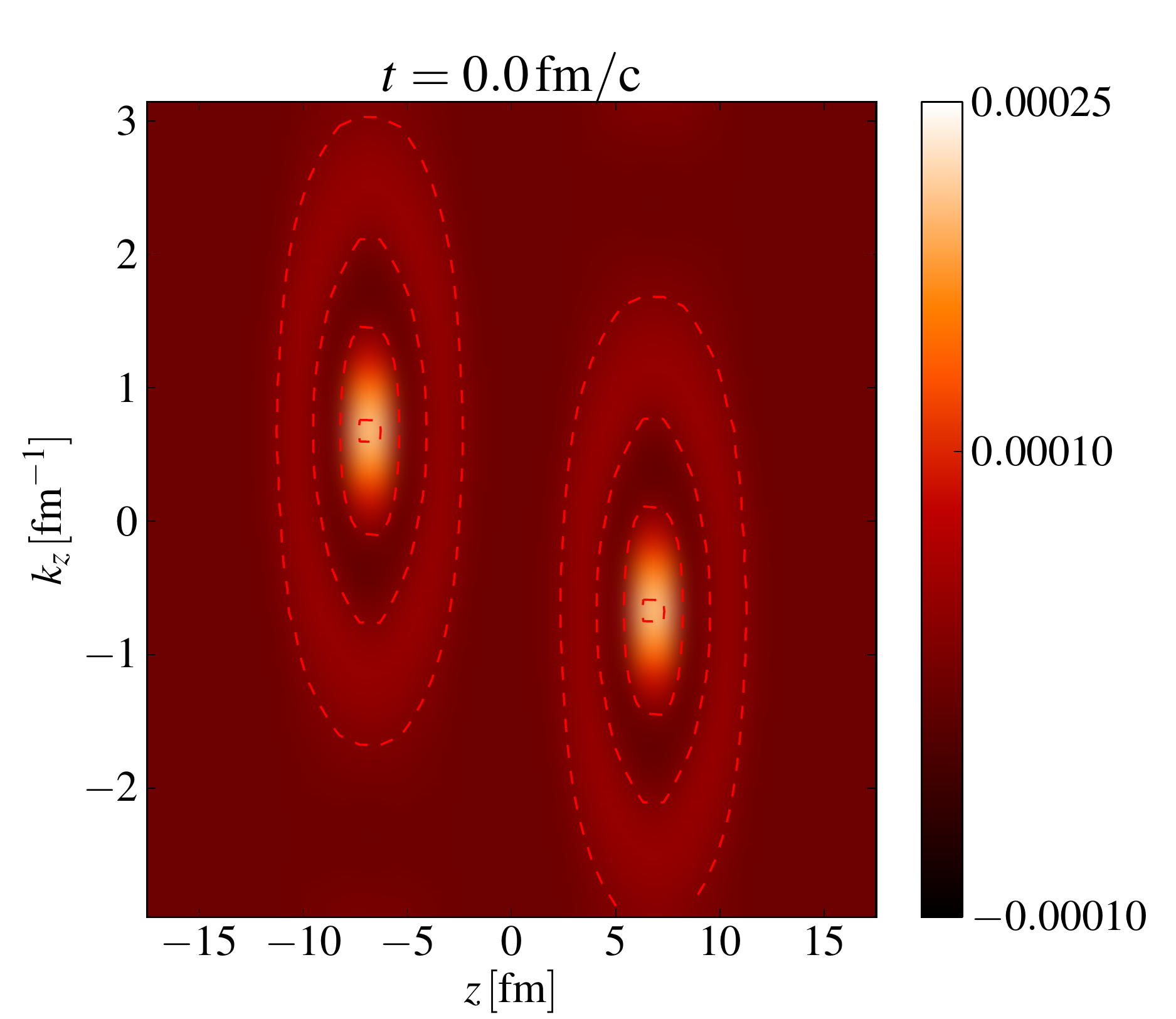}
\setfloatlink{http://th.physik.uni-frankfurt.de/~loebl/vid2.mpeg}
\caption{\label{vid:ca240} (color online)
Same as Video~\ref{vid:ca240} with a center-of-mass energy of
$E_\mathrm{c.m.}=240$\:MeV.}
\end{video}

\subsection{Influence of time-odd terms}

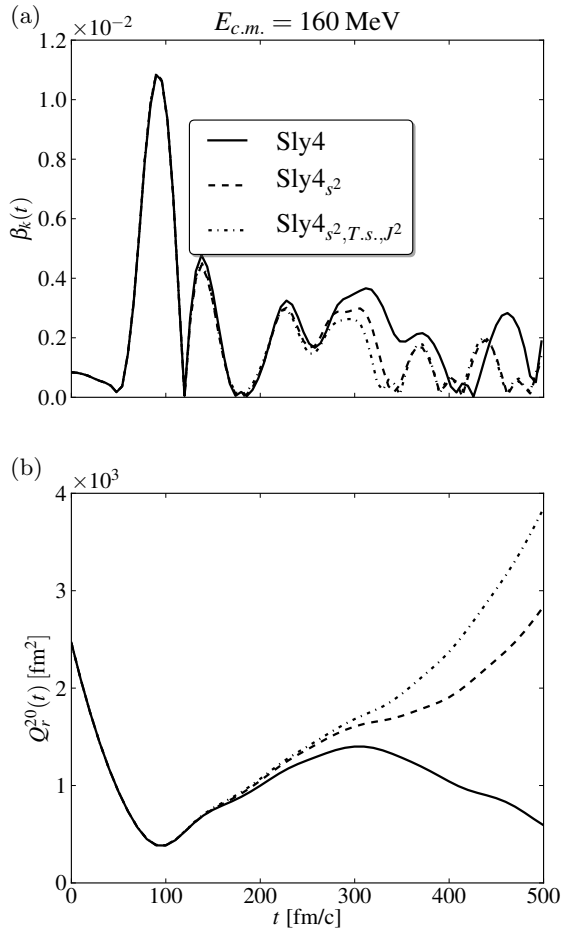
\begin{figure}[hbtp]
 \centering
  \savebox\IBox{\includegraphics[width=8.1cm]{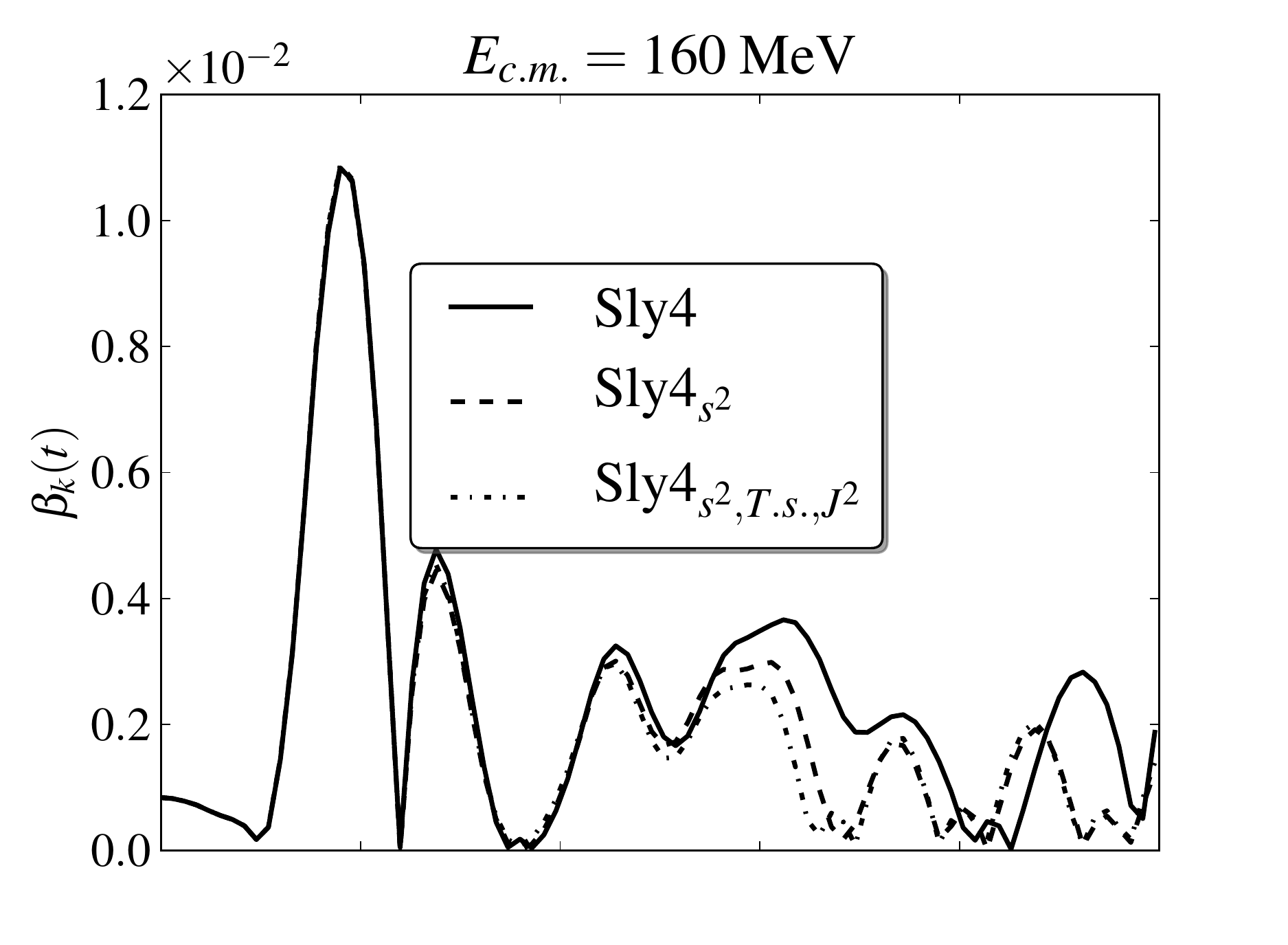}}
 \begin{pspicture}(\wd\IBox,\ht\IBox)
    \rput[lb](-0.5,0.5){\usebox\IBox}
    \rput[lb](-0.3,6.){(a)}  
  \end{pspicture}
  \savebox\IBox{\includegraphics[width=8.1cm]{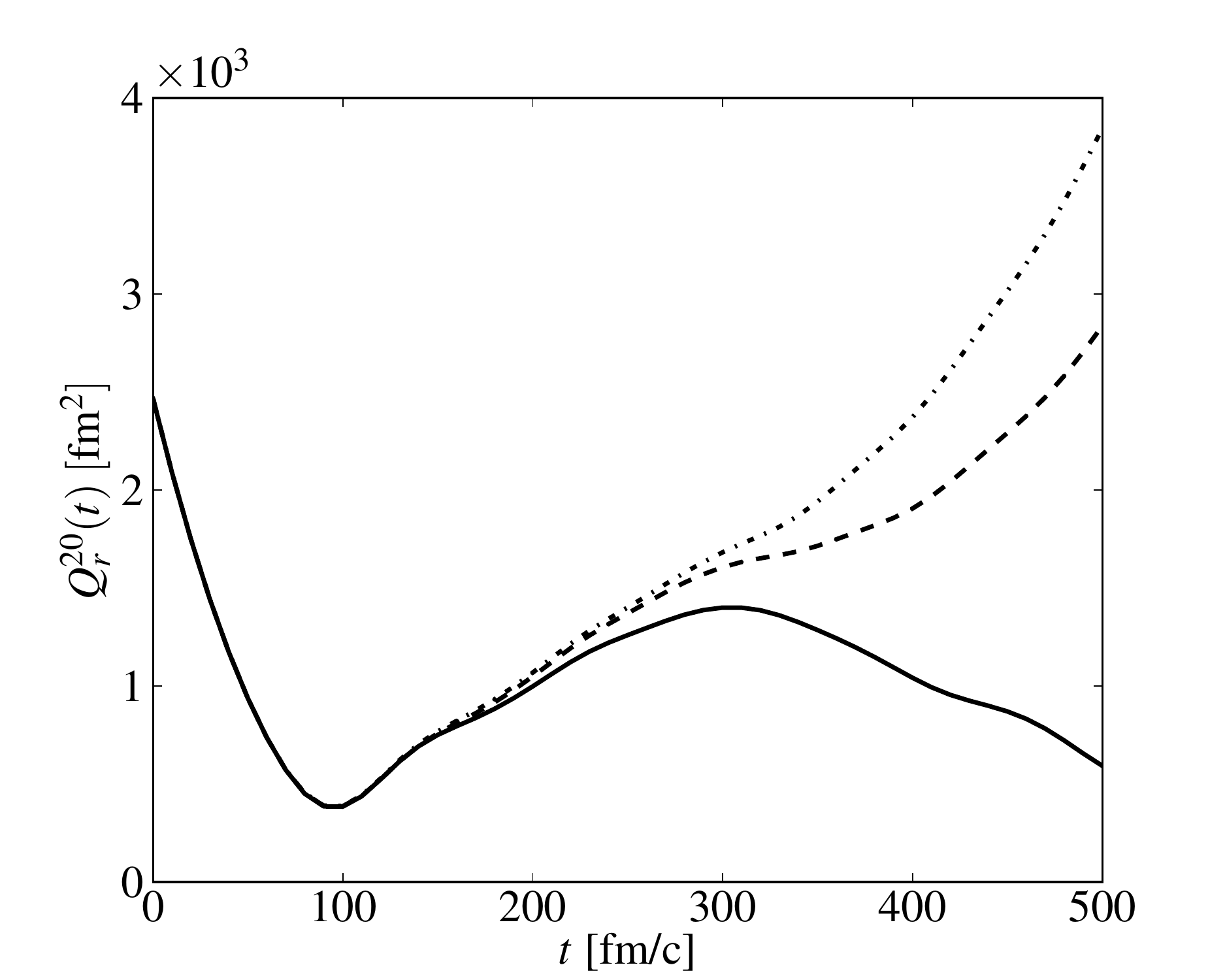}}
 \begin{pspicture}(\wd\IBox,\ht\IBox)
    \rput[lb](-0.5,0.5){\usebox\IBox}
    \rput[lb](-0.3,6.5){(b)} 
  \end{pspicture}
\caption{\label{fig:extra} 
Global observables $\beta_k(t)$ (a), and $Q^{20}_r(t)$ (d) are shown for a central $^{40}$Ca+$^{40}$Ca
collision with fixed Skyrme interaction SLy4.}
\end{figure}

Skyrme energy-density functionals are calibrated to ground state
properties of even-even nuclei~\cite{Chabanat,Bartel,Reinhard2}.
This leaves the choice of the time-odd terms in the functional (current
$\mathbf{j}^2$, spin-density $\mathbf{s}^2$, spin kinetic energy density $\mathbf{T}$,
and the spin-current pseudotensor $\tensor{\mathbf{J}}$) largely unspecified~\cite{BH03}.
Galilean invariance requires at least some of these terms to be present
depending on the presence of the associated time-even term, e.g. $\mathbf{j}^2$
for the $(\rho\tau-\mathbf{j}^2)$ combination.
In our calculations we always include the time-odd part of the spin-orbit interaction.
In order to investigate the effects of the remaining time-odd
terms, we have compared different choices by using a single Skyrme parametrization and the same test case.
We choose the force SLy4 and start with
the minimum number of time-odd terms which is needed to ensure
Galilean invariance~\cite{Eng75a}. In the next stage, we include also
the spin-density terms proportional to $\mathbf{s}^2$. Finally, we also add the combination 
which includes the tensor
spin-current term $(\mathbf{s}\cdot\mathbf{T}-\tensor{\mathbf{J}}^2)$. 
A comprehensive notation of the full Skyrme functional can be found,
e.g. in  Ref.~\cite{Les07}. As shown in Fig.~\ref{fig:extra},
at least for
the quantities $\beta_k$ and $Q^{20}_r$, varying these time-odd terms
has a very little effect in the initial contact phase and the dynamical
behavior becomes somewhat different only in later stages of the collision.
On the other hand, small differences near the threshold energy
(the highest collision energy for a head-on collision that results in a composite system.
At higher energies the nuclei go through each other)
can have large long-term effects on the outcome of the collision. 
For example a small difference in dissipation may be enough to
influence the decision between re-seperation or forming a composite system.
We have also checked a broader range of collision energies
from the fusion regime up to deep inelastic collisions. 
The interesting quantity is the loss of fragment kinetic energy
between the entrance and exit channels. It was found that the spin terms contribute
small changes to this loss which can go in both directions, less
dissipation near fusion threshold and more dissipation above.
Subplot (b) of Fig.~\ref{fig:extra} shows the effects near the Coulomb
barrier where spin terms reduce dissipation.

\section{Summary}
\label{sec:conclusion}

We have presented a geometrically unrestricted framework to study
nuclear dynamics within TDHF in the full six-dimensional phase space.
The impact of different reaction parameters on the outcome of a
heavy-ion collision was studied in detail for the $^{40}$Ca+$^{40}$Ca system.
We find that the occurrence of transparency is clearly reflected 
in the global asymmetry of the Wigner momentum distribution.
The surprising result that in some cases the system merges at higher
energies and shows transparency at lower ones can be related to the 
interplay between momentum- and configuration-space volumes which is a
reflection of the Pauli principle.
It is also interesting that the two distributions in phase-space never
truly combine to form a single distribution. 
This clearly indicates that two-body collisions will be necessary to
achieve true equilibrium as the reaction proceeds to longer contact
times.
The detailed degree of relaxation found depends on energy and also the
properties of the Skyrme force, 
where especially the effective mass seems to be important. The
presence of additional time-odd terms in the Skyrme functional appears
to have
a complex impact on the outcome of a collision as well. 
In this paper only one non-central collision was studied. 
A systematic investigation of impact parameter and energy dependence (fusion, deep-inelastic reactions)
as well as even heavier systems would be highly interesting but is beyond
computational feasibility at the moment.

\section*{Acknowledgment}

This work has been supported by  by BMBF under
contract Nos. 06FY9086 and 06ER142D, and the U.S. Department
of Energy under grant No. DE-FG02-96ER40975 with
Vanderbilt University. The videos linked in the manusscript can be
found at \url{http://th.physik.uni-frankfurt.de/~loebl/vid1.mpeg} and
\url{http://th.physik.uni-frankfurt.de/~loebl/vid2.mpeg}.

\end{document}